# A knowledge-based approach to semi-automatic annotation of multimedia documents via user adaptation

Afzal Ballim, Nastaran Fatemi, Hatem Ghorbel, Vincenzo Pallotta

MEDIA Research Group
DI-LITH, EPFL
IN Ecublens, 1015 Lausanne, Switzerland
{ballim, fatemi, ghorbel, pallotta}@di.epfl.ch

**Abstract**

Current approaches to the annotation process focus on annotation schemas, languages for annotation, or are very application driven. In this paper it is proposed that a more flexible architecture for annotation requires a knowledge component to allow for flexible search and navigation of the annotated material. In particular, it is claimed that a general approach must take into account the needs, competencies, and goals of the producers, annotators, and consumers of the annotated material. We propose that a user-model based approach is, therefore, necessary.

## 1. Introduction

Text annotation is already a rich domain with many divergent points of views concerning the annotation schema to adopt. When we add multimedia or multi-modal information in this context the complexity greatly increases. Such multimedia and multi-modal information already exists in large quantities, so we are already faced with these problems. Standards and guidelines, even those currently proposed, are not in themselves sufficient for a number of reasons. Mechanisms must exist to insure that the application of such standards or guidelines by an annotator is coherent. When faced with multiple annotators this problem is exacerbated. In addition, these schemas are heavily weighted towards the notion of the perspective of the annotator. The producer and the consumer (end-user) of the material are all but forgotten. During the production phase, much material is produced which has an annotation nature. However little is currently available to allow the producer to thus annotate his or her material. In addition, for richly annotated material the needs of the user must be taken into account to prevent him being swamped by information that is not useful to him.

To work towards a resolution of these problems, in this article we propose an approach which federates the producer, the annotator and the consumer in an environment and which reflects the viewpoints of each of them by using intelligent interfaces between the annotations and the users. More specifically, the proposed approach aims at resolving the following problems, which are pertinent to the application area:

- The management of annotations done by different type of users (producer, annotator, and consumer) taking into account the influence of each of them on the others;

- The management of the different levels of users' competences;

- The management of multimedia annotations for multimedia material (the coherence of such annotations and the relations between them);

- The management of multiple annotation schemas.

In this work, we define annotation as any multimedia information that is explicitly added to a "document". This information may have already existed implicitly and have been deduced from the content. It is also possible that the information existed but in another mode (audio, video, etc.). In order to add the annotations we propose a semi-automatic annotation environment that uses automatic modules coordinated by the interaction of different users. A knowledge base containing different user models permits the separation of users' roles in the annotation process. The architecture permits the isolation of the actors and the system is able to adapt its interface according to the actors' roles and their profiles. Users are therefore forced to respect the "design" model, which really corresponds to their competence without limiting their domain of action. A further advantage of such an adaptive architecture is to allow multiple (linguistic, temporal etc.) annotations that can be performed by more than one annotator. A certain degree of confidence is attributed to these parallel annotations according to the actors profiles in order to resolve conflicts if they exist.

## 2. Classical approaches to annotation

Recent research studies in the domain of document annotation have mainly focused on the development of guidelines and standards for annotations. Examples of these are Text Encoding Initiative (TEI), Corpus Encoding Standard (CES), Dublin Core used for text annotation and MPEG-7 which is a current standard being developed for multimedia content description. These guidelines and standards enable the definition of the annotation schema which consists of a terminology (e.g. tag-set, attributes, name space) and a set of rules specifying the application of this terminology. However these specifications do not treat how to annotate in effective annotation systems.

The classical approach of annotation systems is based on a pipeline architecture. First the producer creates the material. Then the professional annotator who is an expert in a specific domain adds a meta-data layer to this material according to an annotation schema based on the domain application requirements. The consumer at the end of this chain exploits the annotated material to accomplish his goals.

In classical annotation systems, the modification and the revision of the annotation schema is either realized in an iterative fashion or by a learning procedure. Generally, the annotation schema should be revised due to two main reasons. First, the users needs may not have been completely considered in design of the annotation schema. Second, the domain application and thus the users' needs may evolve during the time.

### 2.1. Problems with classical annotation

The workflow perspective so far used in the classical approaches draws a large separation between the producer, professional annotator and consumer. This separation generates an over emphasis on the professional annotator and neglects the producer of the material and the consumer of the annotated document in the process of annotation.

During the production phase, much material is produced which has an annotation nature. In current systems, however, the producer is not allowed to participate in the annotation process either explicitly (annotating his material) or implicitly (via a model of his "intentions").

In addition, the consumer viewpoints are generally little considered in the annotations performed by the producer. Similarly, during the annotation, professional annotator's task is limited to the application of the annotation schema to the raw material, without the consideration of other users' viewpoints. When taking into account the user's viewpoints, it is not sufficient to simply consider the user's needs either. It is also necessary to account for the user's knowledge (competencies and roles). The user's knowledge is little considered in most of the previous annotation systems, for instance in the "Alembric" project at MITRE (Day et al., 1997) the generation of annotation rules is based on a training procedure which considers only previous annotations, without really caring about the producers' and consumers' models.

A further drawback of the classical approach is the absence of any account for different levels of users' competence. This can influence the generation and the use of annotations in different manners, for example different levels of confidence shall be assigned to the annotations produced by annotators having different levels of competence. Similarly, different annotations levels shall be produced corresponding to different levels of consumer competence. In addition, for the annotators having the same level of competence, there should exist the possibility of multiple annotation levels. There exist some systems which permit multiple annotation levels, for example the Multilevel Annotation Tool Engineering (MATE) (Carletta et al., 1999) allows the annotation of spoken dialogues with respect to the prosodic, morpho-syntactic, core-reference, dialog acts and communicative aspects levels. MATE system uses this multi-level approach for the visualization of the transcribed text and annotations according to user-defined views.

Of course, if one allows for multiple annotators and multiple levels of annotation then one must address the question of consistency checking i.e. verifying if the annotations done by different annotators (in the same level or different levels) conform to each other.

Standards and guidelines used in current annotation systems provide a tag level specification of terminology. However, the annotation systems need to support the content level specification of terminology. This is typically the main problem in discourse and dialog annotation systems, for instance the Dialog Annotation Tool (DAT) that supports the Dialog Mark-up in several Layers (DAMLS) schema (Core et al., 1997). The latter defines the utterance-tags for annotating dialogue, but leaves open the question of tags for the content of the discourse.

Finally, in classical annotation systems, there is little account for the co-existence of different multimedia annotations, for example the coherence between a video annotation and a sound annotation on the same part of the material. This is becoming an important issue, as in multimedia documents multi-modal interaction is becoming prevalent for non-expert users.

Amongst few research works that share with us the consideration of some of the so-called problems for classical systems, we refer to Dynamic Hyper video (Csinger, 1995). One of the remarkable points in this work is the importance accorded to the "intent" (of the users and the producers) rather than the "content" (of the document). To permit this notion their approach consists of the construction of a user model, which is further used to tailor the presentations of a document to meet the consumer's goals and expectations. However, in this approach, there is no separation between the professional annotators and the producers' roles. Both of these are presented under the concept of the "authors". Consequently there is no consideration of the influence of the producers' model on the annotators' model and vice versa. Moreover there is no account for different levels of competence of the annotators, the coherence of different annotators' annotations, and other points discussed previously.

### 3. A user-centered approach

In this section we propose our approach for dealing with the management of multimedia information added by multiple annotators to the source document in order to obtain a consumable annotated document, and taking into account the needs and competencies of the different users. We outline the notions of a knowledge-based approach for defining the annotators' roles and for representing and assimilating annotations into a suitable annotation repository.

The use of a knowledge base dramatically increases the information which has been added to raw documents by means of simple annotations, because of the amount of implicit information that may be automatically derived from annotations. Moreover such a knowledge base must be carefully designed as it requires the adoption of flexible and efficient management mechanisms. We are going to discuss these issues and to propose some feasible solutions.

### 3.1. Adopt a sound knowledge approach

The adoption of a knowledge-based approach to annotation gives several benefits. First of all, we should distinguish two types of knowledge:

1. Knowledge about users, which:

    - takes into account producers, professional annotators, and consumers;
    - allows different competence levels for each, and reflection on how users of different competence levels can be accounted for, and consider each others.

2. Knowledge on annotations, which requires:

    - the choice of an intermediary KR language for representations that can be used for inference;
    - assimilation of knowledge from annotations.

The first kind of knowledge can be thought as a dynamic taxonomy where users are classified with respect to different dimensions (e.g. needs, competences, access rights, etc.). Since there is no clear and strict separation between possible classes of users (e.g. producers, professional annotators and consumers) we need to define users roles in a more flexible way. We propose to specialize a notion of generic annotator with respect to the above three default roles. In turn we can define new users' classes by multiple inheritance on them giving also the possibility of assigning membership degrees to the respective super-classes. Figure 1 illustrates the above idea.

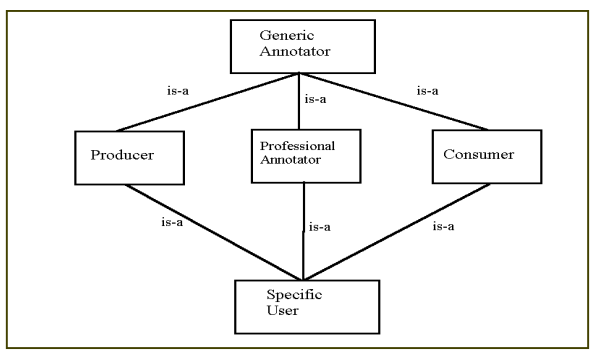

Figure 1: Relations between users' classes

Management of knowledge about annotation requires several important features in order to suitably account for representing document annotations. First of all we need to consider annotation types e.g. linguistic annotation for text analysis, journalistic commentary, etc. Identifying different annotation types will allow us to decide whether various tasks, such as generation of annotations, feature extraction and meta-data creation, can be done automatically, semi-automatically, or manually (but still assisted). To be more precise, annotation types will enforce constraints on the annotation task. These constraints can be of a different nature and they will specify both how the annotator should conform to an annotation schema and how present knowledge could be used for assisting the user during the annotation task. The more logical the nature of these constraints, the more we can use standard reasoning systems for deriving new information and supporting the user by automatic information extraction procedures.

The management of an annotation repository would benefit from having annotations represented using an intermediate logical form. The choice of the logical form is a complex question that cannot be adequately addressed in the confines of this paper. For instance we could store our knowledge base using a deductive database and then exploit well-known knowledge assimilation techniques (Decker, 1998). It is important to observe that the knowledge assimilation process can suitably model the annotation process. In fact, the annotator will have the role of knowledge engineer. He or she will acquire knowledge from the source document and will provide additional information that will be used by consumers for information retrieval and extraction. The additional information will be translated into a logical form and assimilated in the knowledge base and will also serve as semantic indexes for the original document. The consumer will access the document simply by querying the knowledge base and further accessing the indexed document.

The knowledge assimilation for annotation can be outlined, following (Kowalski, 1979), by distinguishing the following four cases when the user adds a new annotation:

1. The logical representation of the annotation can be efficiently deduced from the existing knowledge base and it can be ignored.

2. The logical representation of the annotation extends the knowledge base in a way such that already present annotations can be efficiently derived. In this case the new annotation is inserted and the derived annotations removed.

3. The insertion of the annotation will bring the knowledge base an inconsistent state and therefore must be specially treated.

4. The annotation is logically independent from the current knowledge base and can be inserted.

Efficiency may be a crucial issue since the amount of information may rapidly increase due to the rich semantic content of multimedia documents. In some cases, adding annotations to the knowledge base can be monotonic (it is not necessary to revise and remove previous annotations), but in the general case we cannot rely on this assumption. This means that we need to consider that enabling automatic feature recognition we may be faced with the problem of dealing with (locally and globally) contradictory information.

Adopting a knowledge-based approach for annotation, it is possible to enforce consistency criteria and logically define them as integrity constraints on the database. Due to the presence of multiple annotators and indeed potential parallel annotations, we need to define a notion of annotation consistency in order to avoid the presence of contradictory information. Moreover it seems not feasible to consider only a global notion of consistency because of the problem of efficiency of the resulting system. In general ensuring local consistency by means of prescriptions and recommendations during the knowledge assimilation process may partially avoid the need of global consistency checking.

While there is no consensus on how to represent annotations in a knowledge base nor on how to structure and design it, the use of a deductive database for instance, will allow us to account for temporal information, either at the annotation content level (e.g. adopting for instance event calculus (Kowalski et al., 1995)) or at the annotation history level (e.g. using deductive temporal databases (Sripada, 1988)). The latter possibility will be very useful if we consider multiple annotators for the same document (e.g. for dealing with the problem of versioning).

When using knowledge-based system, reasoning and inference play a central role. In our case, these two features may serve to different purposes: establishing consistency criteria; ascription of users profiles; induction of annotation schema.

One of the main goals of our proposal is about the integration of manual and automatic annotation. Automatic extraction of features from the document may serve as the basis for annotation, but this activity cannot be completely unsupervised. The reason lies in the fact that tools always blindly return information without relying on the current context that is typically held by the annotator. In addition, tools cannot be aware of the consumer goals and thus of specific requirements. However there is still something that can be done automatically. We can consider the extracted features as observations of the phenomena that are embodied in the document. By means of abductive reasoning and relying on a world knowledge base (e.g. an ontology) we can attempt to infer and propose annotations that will be in turn validated and committed upon the annotator.

Assimilation of annotation into a knowledge base can be very useful if we consider the possibility of using it as the basis for induction of users models and annotation schemas. Databases in general offer powerful tools for discovering data correlations and frequent patterns. We can exploit data mining (Holseimer et al., 1994) techniques to extract this kind of information. Moreover we may easily generate aggregate views of several annotated documents belonging to the same class allowing us to induce annotation schema for such document classes. In a similar way it can be possible to aggregate document annotations with respect to other dimensions, like users' roles or target usage.

## 4. Conclusion

In this paper we have presented a user-model and knowledge-based approach to semi-automatic annotation. It is our belief that such an approach will greatly improve the usefulness of annotation systems where the use of the annotations is varied or initially ill defined. In particular, the use of a knowledge base for representing the annotations will allow for the use of richer search techniques, as well as the deployment of inferential search for knowledge discovery. The use of models of different users will allow for the adaptation of the annotations towards actors of different competence levels and with different needs and goals. Current investigations are dealing with an implementation of an annotation architecture that considers the principles we discussed so far.